%% file: main.tex
\documentclass[useAMS,usenatbib]{mnras}

\usepackage[english]{babel}
\usepackage{graphicx,booktabs}
\usepackage{blindtext}
\usepackage{xcolor}
\usepackage{amssymb,amsmath, wasysym}

\newcommand{\mybf}{}

\include{JournalAbbr}    

\begin{document}

\title[Zoomed Simulations in $f(R)$-gravity]
{Zoomed cosmological simulations of Milky Way sized halos in f(R)-gravity}
\author[C. Arnold, V. Springel, E. Puchwein]
{Christian Arnold$^{1,2}$, Volker Springel$^{1,3}$ and Ewald Puchwein$^4$
\\$^1$Heidelberger Institut f{\"u}r Theoretische Studien, Schloss-Wolfsbrunnenweg 35, 69118 Heidelberg, Germany
\\$^2$Institut f{\"u}r Theoretische Physik, Philosophenweg 16, 69120 Heidelberg, Germany
\\$^3$Zentrum f\"ur Astronomie der Universit\"at Heidelberg, Astronomisches Recheninstitut, M\"{o}nchhofstr. 12-14, 69120 Heidelberg, Germany
\\$^4$Kavli Institute for Cosmology Cambridge and Institute of Astronomy, University of Cambridge, \\$\phantom{^4}$Madingley Road, Cambridge CB3 0HA, UK}
\date{\today}
\maketitle

\begin{abstract} 
  We investigate the impact of $f(R)$ modified gravity on the internal properties of Milky~Way sized dark matter halos in a set of cosmological zoom simulations of seven halos from the Aquarius suite, carried out with our code \textsc{mg-gadget} in the Hu \& Sawicki $f(R)$ model. 
Also, we calculate the fifth forces in ideal NFW-halos as well as in our cosmological simulations and compare them against analytic model predictions for the fifth force inside spherical objects. 
We find that these theoretical predictions match the forces in the ideal halos very well, whereas their applicability is somewhat limited for realistic cosmological halos. 
Our simulations show that $f(R)$ gravity significantly affects the dark matter density profile of Milky Way sized objects as well as their circular velocities. 
In unscreened regions, the velocity dispersions are increased by up to $40\%$ with respect to $\Lambda$CDM for viable $f(R)$ models. 
This difference is larger than reported in previous works. 
The Solar circle is fully screened in $\bar{f}_{R0} = -10^{-6}$ models for Milky Way sized halos, while this location is unscreened for slightly less massive objects. 
Within the scope of our limited halo sample size, we do not find a clear dependence of the concentration parameter of dark matter halos on $\bar{f}_{R0}$.  \end{abstract}


\renewcommand{\d}{{\rm d}}

\section{Introduction}
\label{sec:introduction}

The physical origin of the late time accelerated expansion of the Universe is an unsolved and highly debated issue in modern cosmology. Although the standard model of cosmology, the $\Lambda$ cold dark matter ($\Lambda$CDM) model, successfully describes the acceleration and a wide array of cosmological observations, it lacks a compelling explanation for $\Lambda$, motivating the search for possible alternative scenarios.

Such alternative cosmological models can be broadly characterized into two classes {\mybf\cite[see e.g.][]{clifton2012, joyce2016}}. The first class consists of so-called dark energy models which add a new type of field to the energy momentum tensor, and hence modify the source terms in the gravitational field equations. If the field features an equation of state with negative effective pressure, it can account for the accelerated expansion at late times.

The second class of models leaves the source tensor untouched but changes the field equations themselves.  In this work we consider $f(R)$-gravity \citep{buchdahl1970}, which is a representative of these modified gravity models.  Other examples include DGP gravity \citep{dvali2000}, $f(T)$-gravity \citep{bengochea2009} and theories of massive gravity \citep[e.g.][]{hassan2012}.  
These models have in common that they modify the laws of gravity in order to explain the accelerated expansion. They also share the need for some kind of screening mechanism which hides the modifications with respect to general relativity (GR) in our local environment within the Milky Way, otherwise Solar system constraints of gravity that are consistent with GR would be violated. 
Several such screening mechanisms have been explored, including the Chameleon \citep{khoury2004}, the Vainshtein \citep{vainshtein1972, deffayet2002}, the Symmetron \citep{hinterbichler2010} and the Dilaton \citep{gasperini2002} screening.
For $f(R)$-gravity, the chameleon mechanism can ensure GR-like forces in the Solar system \citep{husa2007}.

The non-linearity introduced by the screening mechanism makes numerical simulations essential to fully explore modified gravity theories.  
Numerical works focussing on $f(R)$ gravity have investigated its impact on the matter power spectrum \citep{oyaizu2008, li2012, li2013, llinares2014, puchwein2013, arnold2015}, the mass function of dark matter halos \citep{schmidt2009, ferraro2011, li2011, zhao2011}, cluster concentrations \citep{lombriser2012b} as well as on density profiles \citep{lombriser2012}.  
Further works have investigated the integrated Sachs-Wolfe effect \citep{cai2014}, redshift space distortions \citep{jennings2012}, the properties of voids \citep{zivick2015}, the velocity dispersions of halos \citep{schmidt2010, lam2012, lombriser2012b}, and the properties of semi-analytically modelled galaxy populations \citep{fontanot2013}. 
Recently, hydrodynamical simulations have been used to study galaxy clusters and groups in $f(R)$ gravity \citep{arnold2014}, the Lyman-$\alpha$ forest \citep{arnold2015}, and power spectra and density profiles \citep{hammami2015}. In addition, zoom simulations have been employed to simulate galaxy clusters \citep{moran2014}.

In this work, we for the first time simulate Milky Way sized objects using high-resolution cosmological zoom simulations of $f(R)$-gravity. We employ an upgraded version of our modified gravity simulation code \textsc{mg-gadget} to resimulate a set of seven halos from the Aquarius project \citep{springel2008}. 
Our analysis focuses on the impact of modified gravity on density profiles, gravitational forces, circular velocities as well as velocity dispersions. In addition, we derive an analytic estimate for the $f(R)$-force profile in NFW-halos \citep{navarro1997} and compare this theoretical approximation to the simulation results.

In Section~\ref{sec:fR_gravity}, we introduce $f(R)$-gravity and consider analytical estimates for the modified force. Section~\ref{sec:simulation_code} gives an overview of the simulation code \textsc{mg-gadget} and the performed simulations. Our results are presented in Section~\ref{sec:results}. Finally, we summarize the results and draw our conclusions in Section~\ref{sec:conclusions}.

\section{$\lowercase{f}(R)$-gravity}
\label{sec:fR_gravity}


$f(R)$ models of modified gravity are a generalisation and extension of Einstein's general relativity (GR). In order to account for the accelerated expansion of space at late times, a scalar function $f(R)$ of the Ricci scalar $R$ is added to the action of GR,
\begin{align}
S=\int {\rm d}^4x\, \sqrt{-g} \left[ \frac{R+f(R)}{16\pi G} +\mathcal{L}_m \right],\label{action}
\end{align}
where $G$ is the gravitational constant, $g$ is the determinant of the metric $g_{\mu\nu}$, and the matter Lagrangian is denoted as $\mathcal{L}_m$. A suitable choice of the function $f(R)$ allows eliminating the cosmological constant, which is needed in the standard cosmological model to account for the accelerated expansion.

Carrying out the variation of the action with respect to the metric in the usual way, one obtains the modified Einstein equations \citep{buchdahl1970}, \begin{align} G_{\mu\nu} + f_R R_{\mu\nu}-\left( \frac{f}{2}-\Box f_R\right) g_{\mu\nu} - \nabla_\mu \nabla_\nu f_R = 8\pi G T_{\mu\nu} \label{Eequn}.  \end{align} Here $f_R \equiv {\rm d} f(R)/{\rm d} R$ denotes the derivative of the scalar function with respect to the Ricci scalar, $G_{\mu\nu} = R_{\mu\nu} - \frac{R}{2}g_{\mu\nu}$ is the Einstein tensor, and $T_{\mu\nu}$ is the energy-momentum tensor.  
In the framework of cosmological simulations, i.e.~considering weak fields on scales much smaller than the horizon, one can assume the quasi-static limit and neglect all time derivatives in the above equation {\mybf \citep{oyaizu2008, noller2014, llinares2013, llinares2014b, bose2015}}.  The limitations of this approximation have recently been discussed by \cite{sawicki2015}. 
The field equations then simplify to an equation for the so-called scalar degree of freedom, $f_R$, \citep{husa2007} \begin{align}
 \nabla^2 f_R =  \frac{1}{3}\left(\delta R -8\pi G\delta\rho\right), \label{fRequn}
\end{align}
and a modified Poisson equation,
\begin{align}
 \nabla^2 \Phi = \frac{16\pi G}{3}\delta\rho - \frac{1}{6} \delta R,\label{poisson}
\end{align}
where $\delta R$ and $\delta \rho$ denote perturbations to the background value of the scalar curvature and the matter density, respectively. In order to be consistent with observations, the model must satisfy $|f_R| \ll 1$. To carry out a cosmological simulation in $f(R)$ gravity, one has to numerically solve equations (\ref{fRequn}) and (\ref{poisson}). 
In models with a screening mechanism, equation (\ref{fRequn}) typically involves a highly non-linear dependence on the density field. This is particularly challenging.

\subsection{The Hu \& Sawicki (2007) model} 

All models which modify the laws of gravity should reproduce GR in our local environment in the Milky Way since GR is tested to remarkably high precision in the Solar system. 
For $f(R)$ gravity, this requirement is fulfilled by a class of models featuring the chameleon screening mechanism which suppresses the modifications to GR in high density environments. 
A particularly well studied member of this class is the model proposed by \cite{husa2007}, 
\begin{align}
 f(R) = -m^2\frac{c_1\left(\frac{R}{m^2}\right)^n}{c_2\left(\frac{R}{m^2}\right)^n +1},
\end{align}
where $m^2 \equiv H_0^2\Omega_m$ denotes the mass scale of the model. 
Another requirement for the model is that it closely reproduces the well-tested expansion history of a $\Lambda$CDM universe. This can be achieved by appropriately choosing the three remaining parameters,
\begin{align}
 \frac{c_1}{c_2}=6\frac{\Omega_\Lambda}{\Omega_m} && \text{and}  &&  c_2\left(\frac{R}{m^2} \right)^n \gg 1.\label{eq:c1_c2_lambda}
\end{align}
In our simulations, we adopt $n = 1$.  In the limit $c_2\left(\frac{R}{m^2} \right)^n \gg 1$, one can express the derivative of $f(R)$ in terms of
\begin{align} f_R=-n\frac{c_1\left(\frac{R}{m^2}\right)^{n-1}}{\left[c_2\left(\frac{R}{m^2}\right)^n+1\right]^2}\approx-n\frac{c_1}{c_2^2}\left(\frac{m^2}{R}\right)^{n+1}.\label{fR}
\end{align}

Let us now replace the two parameters of the model, $c_1$ and $c_2$, by a more natural choice. The background curvature of a Friedman-Robertson-Walker universe is given by
\begin{align}
 \bar{R}=12 H^2 + 6\frac{\text{d}H}{\text{d}\ln a}H.
\end{align}
For a flat $\Lambda$CDM expansion history, this simplifies to 
\begin{align}
  \bar{R}=3m^2\left[ a^{-3} + 4\frac{\Omega_\Lambda}{\Omega_m} \right]. 
\end{align}
At $a=1$, the two parameters $c_1$ and $c_2$ are now fully constrained by fixing $\Omega_\Lambda$, $\Omega_m$, $H_0$, $n$, as well as the background value of the scalar field $f_{R0} \equiv f_R(z=0)$. In the following we will adopt $f_{R0}$ as the parameter specifying the Hu \& Sawicki $f(R)$-gravity model. 

\subsection{The fifth force in a spherical overdensity}
\label{subsec:spherical}
Given a complex density distribution, it is in general not possible to solve the above equations of motion analytically.  Nevertheless, one can calculate an analytical estimate for the fifth force for a simple spherically symmetric problem \citep[following][]{davis2012, sakstein2013, vikram2014}.
Consider a spherical overdensity of radius  $R$ and density profile $\rho(r)$ which is embedded in a homogeneous  background density $\rho_0$. If at least part of the object is screened, there will be some screening radius $r_s$ inside which the $f(R)$ modifications to gravity are completely suppressed \citep{davis2012}.  
When approaching $r_s$ from the outside, the ratio of the fifth force to the GR force will monotonically drop from its background value to zero. The cases $r_s \geq R$ and $r_s = 0$ refer to the fully screened and unscreened situations, respectively.
\begin{figure}
\centerline{\includegraphics[width=\linewidth]{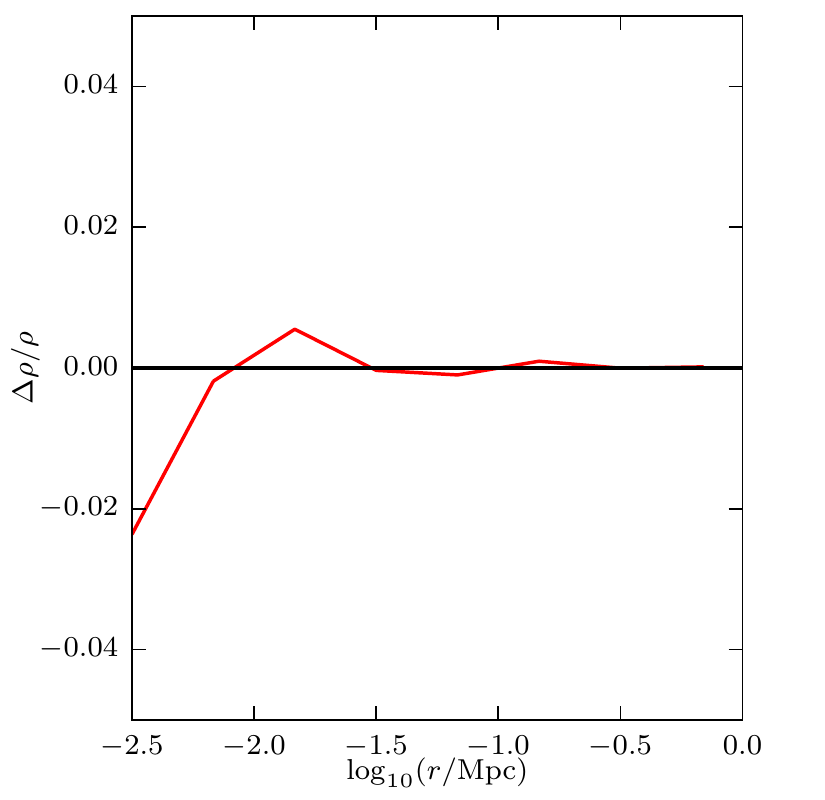}} 
\caption{\mybf{Convergence test for the time integration scheme: Relative difference between the density profiles of a B halo simulated with the standard modified gravity timestep and with a four times smaller one. Both simulations were carried out for the F6 model.}}
\label{fig:convergence}
\end{figure}

\begin{figure}
\centerline{\includegraphics[width=\linewidth]{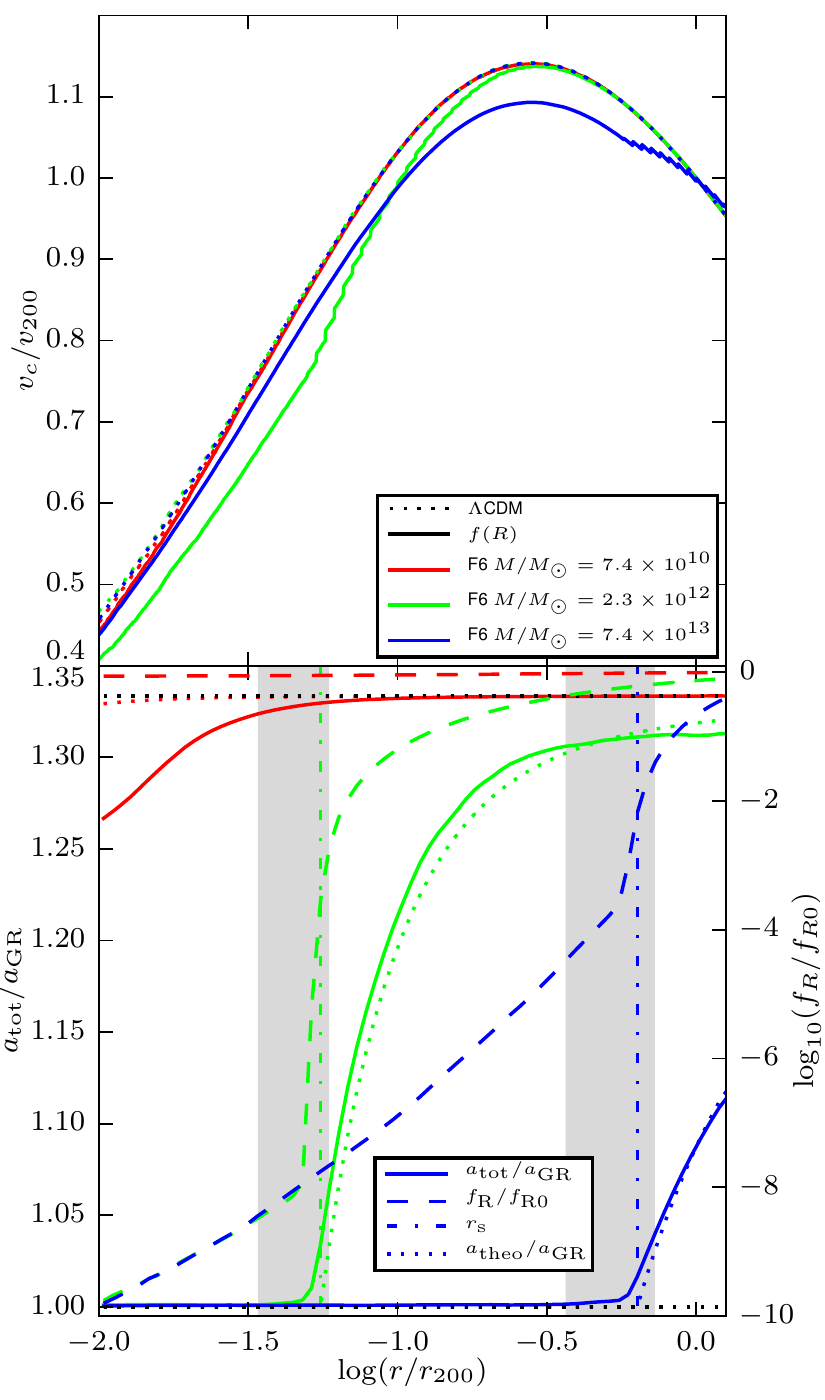}} 
\caption{\textit{Upper panel:} The circular velocity profiles for ideal NFW halos of three different masses but equal concentration $c\approx 10$ for $f_{R0} = -10^{-6}$ (solid lines) and $\Lambda$CDM (dotted line) scaled with the circular velocity at $r_{200}$. 
The velocities are derived from the enclosed masses, taking into account the increased gravitational forces for $f(R)$ gravity in unscreened and partially screened regions. \textit{Lower panel:} The solid lines show the ratio of total acceleration to GR acceleration for the three different halos in $f_{R0} = -10^{-6}$ cosmology. 
Dotted lines show the theoretical expectations for this force ratio. The corresponding values of the scalar field are plotted as dashed lines. The theoretical values for the radius at which we expect screening to set in (obtained from Eqn.~\ref{theo_rs}), $r_s$, are shown as dashed-dotted lines for the heavy and the intermediate mass halo. 
For the least massive object, this radius is zero. The grey shaded regions show an estimate for the uncertainty of this radius. 
The highest and lowest allowed values for $a_{\rm tot}/a_{\rm GR}$ of $4/3$ and $1$, respectively, are indicated by the black dashed lines.}
\label{fig:NFW}
\end{figure}
\begin{figure}
\centerline{\includegraphics[width=\linewidth]{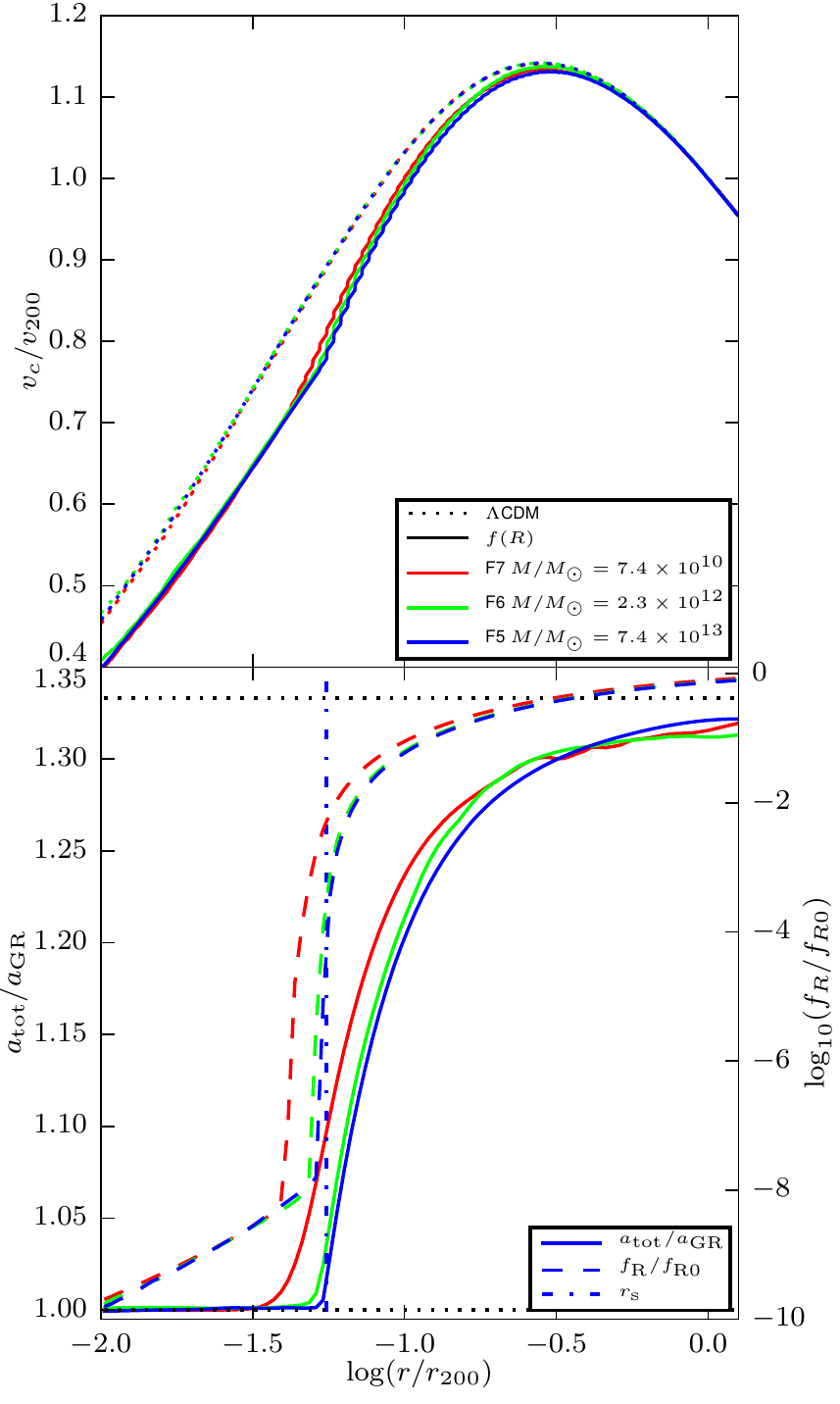}}
\caption{Same as Figure~\ref{fig:NFW}, but with the masses and the background values of the scalar field $f_{R0}$ scaled such that the screening radius $r_s$ in units of $r_{200}$ is constant for all three halos. }
\label{fig:NFW_selfsim}
\end{figure}

To estimate the fifth force due to $f(R)$ (i.e.~the excess force relative to Newtonian gravity), 
let us first define a field $\phi$ via
\begin{align}
{\rm e}^{-\frac{2 \beta \phi}{M_{\rm pl}}} =  f_R + 1, \label{def_phi}
\end{align}
where $b = \sqrt{1/6} $ for the $f(R)$-models of interest \citep{brax2008}, and apply the conformal transformation 
\begin{align}
\tilde{g}_{\mu\nu} = {\rm e}^{-\frac{2 \beta \phi}{M_{\rm pl}}} {g}_{\mu\nu}.
\end{align}
The action (\ref{action}) then becomes \citep{brax2008}
\begin{align}
S = \int {\rm d}^4x \sqrt{-\tilde{g}} \left[ \frac{M_{\rm pl}^2}{2}\tilde{R} - \frac{1}{2}\tilde{g}^{\mu\nu} \nabla_\mu \phi \nabla_\nu \phi -V(\phi) +\tilde{\mathcal{L}}_m \right] ,
\end{align} 
where
\begin{align}
V(\phi) = \frac{M_{\rm pl}^2 [R f_R - f(R) ]}{2 (f_R + 1) ^2}.
\end{align}
$\tilde{R}$ is the Ricci scalar corresponding to the metric $\tilde{g}_{\mu\nu}$.

In the Newtonian limit, the field equations for $\phi$ can be written as
\begin{align}
\nabla^2 \phi = \frac{\partial V}{\partial \phi} + \frac{\beta \rho}{M_{\rm pl}}. \label{phi_field}
\end{align}
One now has to distinguish different cases. If the object is at least partially screened, the effective potential will reach its minimum inside $r_s$ and we have \citep{hui2009}
\begin{align}
 \frac{\partial V}{\partial \phi}  =  -\frac{\beta \rho}{M_{\rm pl}}.
\end{align}
In other words, the derivative of the field $\phi$ will be constant inside $r_s$, and since there are no sources, $\phi = \text{const.}$ Far outside the sphere (for $r \gg R$), the field $\phi_0$ is just given by the background value $f_{R0}$ of the scalar degree of freedom. 
To obtain $\phi$ in the remaining region in between, i.e.~in the partially screened shell of the sphere, one can linearize Eqn.~(\ref{phi_field}) and express it in terms of perturbations of the background value $\delta\phi = \phi - \phi_0$,
\begin{align}
\nabla^2 \delta\phi = \frac{\partial^2 V}{\partial \phi^2}  \delta\phi + \frac{\beta\, \delta\rho}{M_{\rm pl}}. \label{phi_rel}
\end{align}
Writing the density in Eqn.~(\ref{phi_rel}) in terms of the Newtonian potential $\nabla^2 \Phi_N = 4 \pi G \rho$, integrating twice, and resubstituting the Newtonian potential for a spherical overdensity, ${ \d \Phi_N}/{\d r} = {G M(<r)}/{r^2}$, one arrives at an expression for the fifth force for $r>r_s$ \citep{sakstein2013, davis2012}:
\begin{align}
F_{\rm modgrav} = \alpha \frac{G M(<r)}{r^2} \left[ 1 - \frac{M(r_s)}{M(<r)} \right],\label{theo_ff}
\end{align}
where $\alpha = 2 \beta^2 = 1/3$ is the coupling strength of $f(R)$ gravity.  

What remains to be done in order to obtain the fifth force is to estimate the screening radius $r_s$. It is implicitly given by the integral equation \citep{sakstein2013}
\begin{align}
\frac{\phi_0}{2 \beta M_{\rm pl}} = 4 \pi G \int\limits^R_{r_s} r\, \rho(r)\, \d r .\label{theo_rs}
\end{align} 
Equations (\ref{theo_ff}) and (\ref{theo_rs}) yield an estimate for the radius inside which the object is fully screened as well as the fifth force profile for objects which are roughly spherical (as the dark matter halos we simulate in this work). 
Given the density profile of a simulated halo, one can easily compute an approximate estimate for the fifth force and compare to the simulation outcomes. 
The only remaining question is which radius one should choose for the outer boundary $R$, as in practice it is hard to judge where an halo exactly ends. In this work, we use $r_{200}$ (the radius which encloses a sphere with a mean density of $200$ times the critical density) as a natural choice for $R$.

Let us now assume that the density of the halo is given by a NFW-profile \citep{navarro1997}:
\begin{align}
\rho(r) = \frac{\rho_{\rm crit} \delta_c}{\left(\frac{r}{r_{\rm\tiny NFW}}\right) \left( 1 + \frac{r}{r_{\rm\tiny NFW}}\right)^2}.\label{nfw}
\end{align}
To avoid confusion with the screening radius $r_s$ we denote the scaling radius of the NFW-profile as $r_{\rm NFW}$ here. Inserting Eqn.~(\ref{nfw}) into (\ref{theo_rs}), and solving for $r_s$ gives
\begin{align}
r_s = \frac{r_{\rm NFW}}{\frac{1}{1+r_{200}/r_{\rm NFW}} - \frac{3 \ln(f_{\rm R0} + 1)}{8 \pi G \rho_{\rm crit} \delta_c r_{\rm NFW}^2}} - r_{\rm NFW}.\label{rs_nfw}
\end{align}

The scale introduced by the screening radius will obviously break the self-similarity of dark matter halos with equal concentration but different masses as known in the standard model of cosmology. Scaling both halo mass and $f_{R0}$ such that the ratio $r_{s}/r_{200}$ and the concentration parameter stay constant for different $f(R)$ models is nevertheless possible. 
This restores some kind of self-similarity in $f(R)$ gravity:
\begin{align}
\left(\frac{M_1}{M_2}\right)^{\frac{2}{3}} = \frac{\ln(f_{R0_1}+1)}{\ln(f_{R0_2}+1)} \approx \frac{f_{R0_1}}{f_{R0_2}},\label{scaling}
\end{align}
where $M$ denotes $M_{200}$ {\mybf and the subscripts 1 and 2 refer to the first and the second model/halo, respectively.}
As a cautionary remark it is important to say that this involves a scaling of $f_{R0}$ and will therefore not work for a given fixed $f(R)$ model.

\section{Simulations and methods}
\label{sec:simulation_code}
Using the same initial conditions as the Aquarius project \citep{springel2008, marinacci2014} we carry out for the first time zoom simulations in $f(R)$-gravity of a set of 7 Milky Way sized halos (A, B, C, D, E, G and H in the \citealt{marinacci2014} terminology) employing the cosmological simulation code \textsc{Modified Gravity gadget} (\textsc{mg-gadget}, \citealt{puchwein2013}). 
For all halos we simulate the evolution of the matter distribution for $\bar{f}_{R0}=-10^{-6}$ (referred to as F6), $\bar{f}_{R0}=-10^{-7}$ (F7), and for the $\Lambda$CDM cosmology as a reference. 
We use $\Omega_m = 0.25$, $\Omega_\Lambda = 0.75$, $h_0 = 0.73$ as our set of primary cosmological parameters. The mass resolution in the zoomed region reaches $3.14\times 10^6\, {\rm M}_\odot$.

\textsc{mg-gadget} is a modified cosmological simulation code based on \textsc{p-gadget3}, which in turn has its origin in \textsc{gadget2} \citep{Springel2005c}. 
It is currently capable of performing simulations of the \cite{husa2007} $f(R)$-gravity model, both for collisionless and hydrodynamical simulations. 
We note that \textsc{mg-gadget} has recently been tested against other $f(R)$ simulation codes and was found to produce comparable results \citep{winther2015}.
 For the present work, we upgraded the modified gravity solver so that it can be used efficiently for zoom simulations in $f(R)$ gravity as well {\mybf \citep[see][for other ways to speed up modified gravity simulation codes]{Winther2014, barreira2015}}. 
 In the following, we give a brief overview of the inner workings of \textsc{mg-gadget}, focussing on the changes that were necessary for zoom simulations. A more comprehensive description of the code can be found in \citet{puchwein2013}.

To solve Eqn.~(\ref{fRequn}) for the scalar degree of freedom, the code employs an iterative Newton-Gauss-Seidel scheme. The iterations are carried out on an adaptively refined mesh (AMR mesh) which allows for higher resolution in high-density regions, and in particular in the zoom region of the simulation box. 
Since the refinement criterion is based on the particle number in a mesh cell, there is no need to modify the algorithm for zoom simulations, except for the performance optimisations discussed below. 
The advantage of this iterative method is that it can solve the highly non-linear equations in a computationally efficient way.  
To ensure that the value of the scalar field $f_R$ stays strictly negative (unphysical positive values would prevent the code from continuing the iteration), \textsc{mg-gadget} does not solve for $f_R$ directly but for $u \equiv \ln [f_R /\bar{f}_R(a)]$ \citep[a method first introduced by][]{oyaizu2008}.

Knowing the value of the scalar field, one can rewrite the modified Poisson equation in terms of an effective density 
\begin{align}
\delta\rho_{\rm eff}= \frac{1}{3} \delta\rho - \frac{1}{24\pi G}\delta R \label{rhoeff},
\end{align}
which accounts for all $f(R)$ effects on gravity including the chameleon mechanism. The modified Poisson equation then reads
\begin{align}
\nabla^2 \Phi = 4\pi G (\delta\rho + \delta\rho_{\rm eff}).\label{rho}
\end{align}
The curvature perturbation $\delta R$ is obtained from
\begin{align}
\delta R = \bar{R}(a)\left( \sqrt{\frac{\bar{f}_R (a)}{f_R}}-1  \right).\label{deltaR}
\end{align} 
By adding the effective density to the real mass density it is thus possible to compute the gravitational forces employing \textsc{p-gadget3}'s TreePM gravity solver. In runs with hydrodynamics, the hydrodynamical forces can be computed using an entropy conserving smoothed particle hydrodynamics scheme \citep{springel2002} as already included in \textsc{p-gadget3}.

In previous versions of \textsc{mg-gadget}, all force computations
were carried out on the same timestep. This is rather time consuming,
especially for zoom simulations which span a wide dynamic range of
timescales.  
{\mybf We therefore employed the same operator split approach which is used in the TreePM force calculation scheme of} \textsc{p-gadget3}{\mybf .
In the standard version of this method, the PM-force is only calculated on (comparatively large) global timesteps, while the timestep for the tree-force is individually adapted for each particle based on an acceleration criterion \citep[see][for a more detailed description]{Springel2005c}.
For }\textsc{mg-gadget,} {\mybf we now couple the calculation of the modified gravity forces to the global PM-timestep.
In order to avoid loss of precision in the fifth force calculation, the criterion for the global timesteps was adapted as well.
In the new method, the global modgrav-PM timestep size, $\Delta t_{\rm modgrav\, PM} $, is given by
\begin{align}
\Delta t_{\rm modgrav\, PM} = \min \left( \Delta t_{\rm modgrav}^{\rm global}, \Delta t_{\rm PM}\right),\label{MG_PM_timestep}
\end{align}
where $\Delta t_{\rm modgrav}^{\rm global}$ is the global modified gravity timestep and $\Delta t_{\rm PM}$ is the standard PM timestep.
The modified gravity timestep is in turn obtained from an acceleration criterion which is similar to the one used for the tree-forces in }\textsc{p-gadget3}, {\mybf 
\begin{align}
\Delta t_{\rm modgrav}^{\rm global}  &= \min\limits_{\rm particles} \left( \sqrt{f  \times l^{\rm i}_{\rm soft}/ a^{\rm i}_{\rm modgrav}} \right).\label{MG_timestep}
\end{align}
$l^{\rm i}_{\rm soft}$ and $a^{\rm i}_{\rm modgrav}$ denote the softening length and the fifth force acceleration for particle $i$, respectively. The prefactor $f$ depends on the integration accuracy parameter.}

An important advantage of this scheme is that
regions which demand very small time-steps (such as the interiors of
galaxies or galaxy clusters) are often screened.  For many
cosmological setups, the permissable modified gravity timestep will
thus be orders of magnitude larger than the timestep required for
standard gravity, making this method particularly effective.

{\mybf To ensure that the above time integration scheme converges for zoom simulations, we performed a convergence test for the B halo in the F6 cosmological model, comparing the above configuration with a setup with four times smaller modgrav timestep: $\Delta \tilde{t}_{\rm modgrav} = \frac{1}{4} \Delta t_{\rm modgrav}$.
We find that density profiles, velocity dispersion and acceleration profiles agree at the $2\%$-level for $r > 10^{-2}\, r_{200}$. Figure~\ref{fig:convergence} shows the relative difference in the density profile between the two runs.}

Our code \textsc{mg-gadget} also includes an inlined version of the \textsc{subfind} algorithm \citep{Springel2001}, which we use to identify gravitationally bound halos and subhalos. 
We identify the centers of halos as the minimum of the gravitational potential. Besides the standard outputs of an N-body code (particle positions, masses, velocities, GR-gravity accelerations) we also include in the output the modified gravity acceleration and the scalar field itself, interpolated from the mesh points of the AMR grid to the particle positions.

\section{Results}
\label{sec:results}
\subsection{Ideal NFW halos}
\label{subsec:NFW}
\begin{figure}
\centerline{\includegraphics[width=\linewidth]{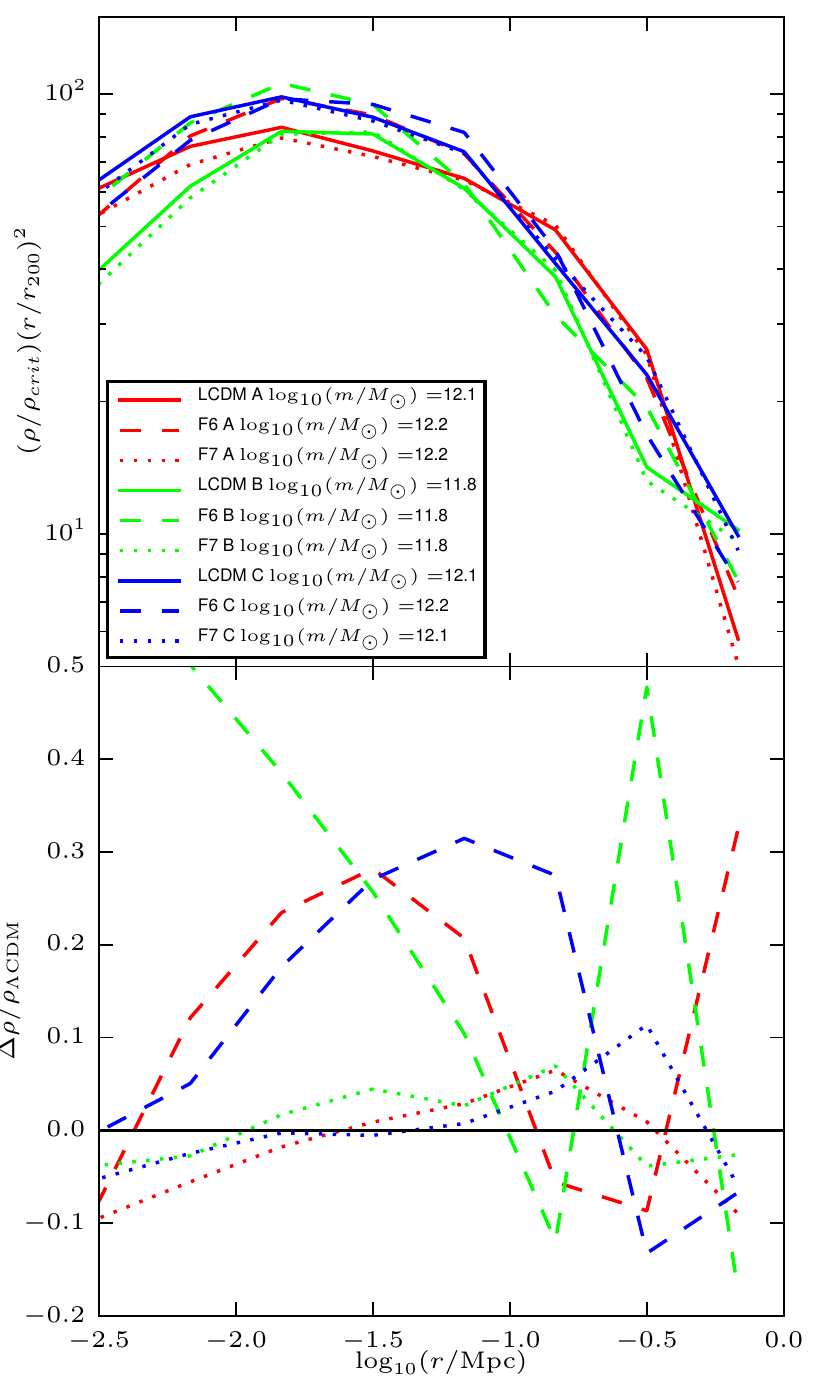}}
\caption{Density profiles of the Aquarius halos A (red lines), B (green lines) and C (blue lines) for the three cosmological models $\Lambda$CDM (solid lines), F6 (dashed lines) and F7 (dotted lines). 
\textit{Upper panel:} The density relative to the critical density multiplied by $(r/r_{200})^2$. 
\textit{Lower panel:} Relative difference between the densities in $f(R)$ cosmology and the corresponding $\Lambda$CDM values {\mybf (this is not identical to the relative differences in the upper panel, since no scaling with $(r/r_{200})^2$ has been applied here)}. The solid black line indicates equality.}  
\label{fig:aqua_density}
\end{figure}
\begin{figure}
\centerline{\includegraphics[width=\linewidth]{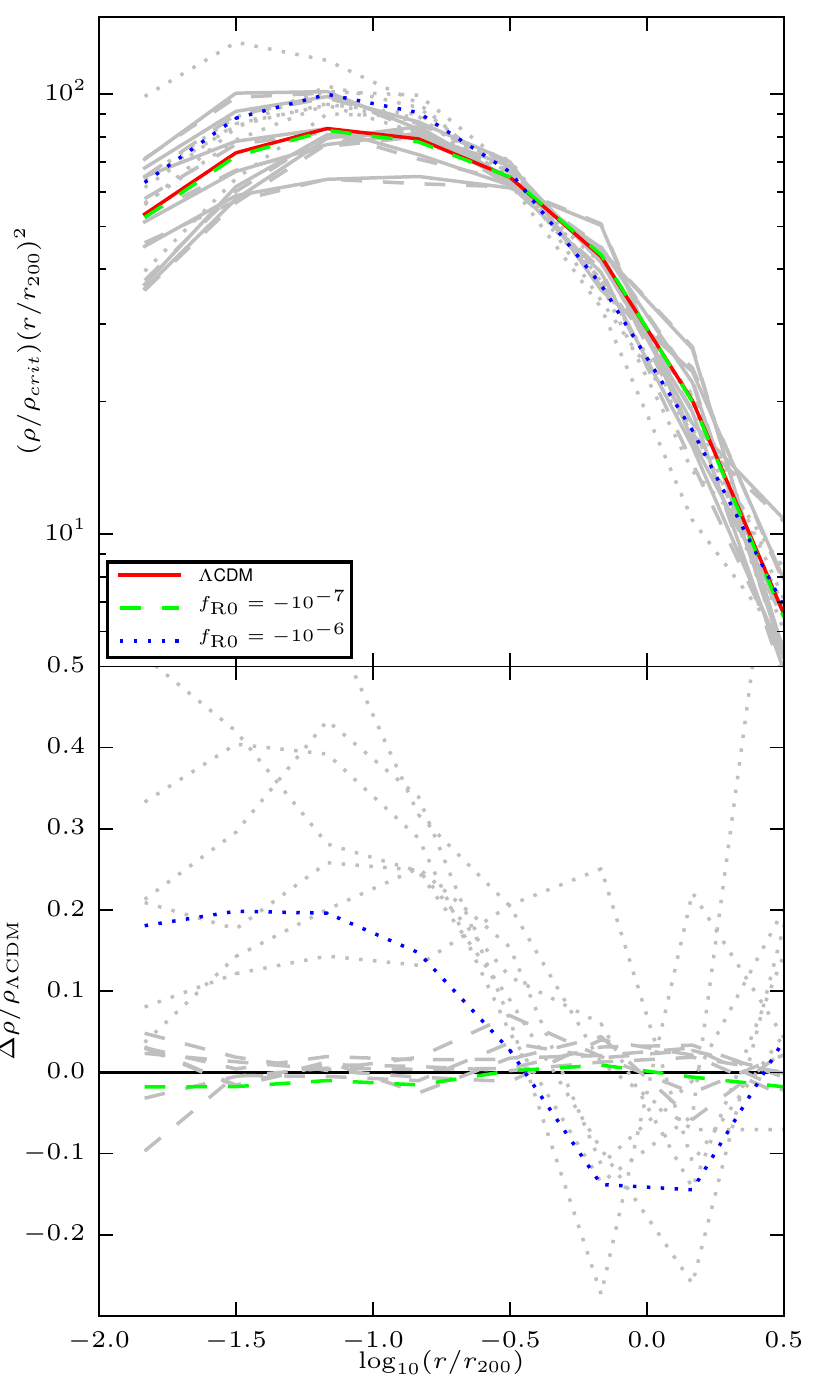}}
\caption{\textit{Upper panel:} Stacked density profiles for all simulated Aquarius halos for $\Lambda$CDM (red solid line), F6 (green dashed line) and F7 (blue dotted line) cosmology. 
\textit{Lower panel:} Relative difference of the stacked density profiles to the $\Lambda$CDM values. The grey lines indicate the densities of the individual halos. {\mybf Again, this quantity is not equal to the relative differences of the values in the upper panel as the latter were scaled with $(r/r_{200})^2$}.}
\label{fig:aqua_density_stacked}
\end{figure}
The theoretical estimates derived in section~\ref{subsec:spherical} assume perfect spherical symmetry. This is of course not true for the simulated halos from the Aquarius suite. To cross-compare the accuracy of the simulations and the theoretical approximations in a more controlled environment first, we set up initial conditions for a collection of three perfectly symmetric halos with a NFW density profile. 
The halos have equal concentration of $c\approx10$ but different mass. Each object is situated in a cubic box of $100\,{\rm Mpc}$ side-length and constant background density.  
The halos serve as initial conditions for \textsc{mg-gadget} to obtain circular velocity profiles, accelerations and $f_R$-profiles based on the code's multi-grid $f(R)$ solver.  

Figure \ref{fig:NFW} displays the results of these tests. The upper panel shows the circular velocity profile in units of $v_{200} \equiv \sqrt{G M_{200} / r_{200}}$ for $f_{R0}=-10^{-6}$ as well as for a $\Lambda$CDM reference simulation for each of the halos.  
Our values for the velocity profiles are obtained from the enclosed mass but with an additional boost accounting for the -- in unscreened regions -- higher accelerations in $f(R)$-gravity $v_c = \sqrt{G M / r}\times \sqrt{a_{\rm tot} / a_{\rm GR}}$.  
In $\Lambda$CDM, the velocity profiles for the three halos overlap almost perfectly, which is expected due to the self-similarity between halos of equal concentrations.  
This self-similarity is broken in modified gravity because of the scale introduced by chameleon screening.

If the object is massive enough, the gravitational potential will drop below a certain threshold at the screening radius $r_s$, causing the chameleon screening to set in. 
As a result, the fifth force quickly decreases to zero.  This is exactly what one can see in Figure~\ref{fig:NFW}. The circular velocity profiles in the upper panel do not coincide anymore in $f(R)$ cosmology.  
For the two more massive objects, there is a tilt in the velocity curves at a certain radius depending on the mass of the object causing the circular velocities to drop with increased screening. 
Having a look at the lower panel, this can be easily explained by the force ratio of total-to-GR force. For the least massive halo, the force ratio stays roughly constant at the theoretically expected value of $4/3$ (indicated by the black dotted line) because even in the center the gravitational potential is not deep enough to trigger screening. 
The slight deviations at small radii are due to the lack of resolution in the AMR grid of the multigrid solver (the size of the grid cells is of the order $10^{-2} r_{200}$ for this object). 

The force ratio of the intermediate mass object is very close to the theoretically expected value for unscreened regions in the outer part as well. But moving inwards, the ratio starts to decrease and quickly drops to unity. 
The radius at which the fifth force becomes negligible is almost exactly at the theoretically predicted value for $r_s$, which was calculated from Eqn.~(\ref{theo_rs}). The grey shaded regions indicate the uncertainty range of this radius.  The errors were obtained by varying the outer integration bound $R$ in (\ref{theo_rs}) from $r_{200}/2$ to $2\, r_{200}$.
Comparing the force ratio of the simulation with the theoretical estimate calculated from Eqn.~(\ref{theo_ff}) shows remarkably good agreement, too.  
The largest halo is already partially screened at the outermost radius shown in Figure~\ref{fig:NFW}. The gravitational potential well of the object is so deep that it crosses the screening threshold already in the outskirts of the halo. Again, both the screening radius and the force ratio show a high level of agreement with the theoretical expectations.  
From Figure~\ref{fig:NFW}, one can also see that the value of the scalar field drops by several orders of magnitude at the screening radius, underlining its highly nonlinear behaviour.

Next, we investigate if the self similarity of the DM halos in the $\Lambda$CDM cosmology can be restored in $f(R)$-gravity by a suitable rescaling of the background field amplitude $f_{R0}$. To this end, we scale $f_{R0}$ according to Eqn.~(\ref{scaling}) such that the ratio $r_s/r_{200}$ of the high and low mass halos are the same as for the intermediate mass object. We also use the same concentration.
Figure~\ref{fig:NFW_selfsim} displays the circular velocity profiles and the total-to-GR force ratio for the objects. In contrast to the previous plot, a good agreement of the $f(R)$ circular velocities can be observed. 
The force ratios and $f_R$ profiles are very similar as well. Only the lowest mass halo shows a slight deviation from the others which can again be explained by the worse resolution of the AMR-grid relative to the halo size in the center of the object. 
The screening radius is -- by construction -- exactly the same.  Knowing the impact of $f(R)$ modified gravity on a certain property for a given value of $f_{R0}$, it is thus possible to predict how the property would change for a different $f_{R0}$ by scaling all masses according to Eqn.~(\ref{scaling}).

\subsection{The Aquarius halos}
\begin{table}\centering
    \begin{tabular}{ l c c c c r }\toprule & $M_{200}$ & ${\mybf r_{200}}$ &  $V_{\rm max}$ & $r_{\rm max}$ & $c\phantom{_{m}}$ \\ 
    & $[10^{12} M_\odot]$ & $[{\rm\mybf kpc}]$ & $[{\rm km} / {\rm s}]$ & $[{\rm kpc}]$  &  \\ 
    \midrule GR A & 1.846 & \mybf 246.1 & 209.13 & 30.46 & 15.24 \\ F7 A & 1.954 & \mybf 250.8 & 206.52 & 40.79 & 11.95 \\ F6 A & 2.020 & \mybf 253.6 & 229.19 & 41.87 & 12.72 \\ 
    \midrule GR B & 0.821 & \mybf 187.8 & 158.62 & 43.72 & 9.10  \\ F7 B & 0.863 & \mybf 191.0 & 160.54 & 42.78 & 9.36 \\  F6 B & 0.919 & \mybf 195.0 & 182.22 & 38.92 & 11.22 \\ 
    \midrule GR C & 1.772 & \mybf 242.7 & 223.07 & 33.76 & 14.78 \\ F7 C & 1.811 & \mybf 244.5 & 222.08 & 32.96 & 15.01 \\ F6 C & 2.294 & \mybf 264.6 & 241.75 & 48.01 & 11.89 \\ 
    \midrule GR D & 1.800 & \mybf 244.0 & 204.78 & 57.43 & 8.97  \\ F7 D & 1.871 & \mybf 247.2 & 206.28 & 56.46 & 9.16 \\  F6 D & 2.251 & \mybf 262.9 & 224.31 & 56.70 & 9.78  \\ 
    \midrule GR E & 1.192 & \mybf 212.7 & 179.95 & 57.26 & 8.08  \\ F7 E & 1.229 & \mybf 214.9 & 183.08 & 58.94 & 8.00 \\  F6 E & 1.324 & \mybf 220.3 & 205.66 & 42.38 & 11.55 \\ 
    \midrule GR G & 1.034 & \mybf 202.9 & 154.61 & 82.35 & 5.21  \\ F7 G & 1.077 & \mybf 205.6 & 154.02 & 60.11 & 6.81 \\  F6 G & 0.984 & \mybf 199.5 & 179.41 & 34.46 & 12.22 \\ 
    \midrule GR H & 0.852 & \mybf 190.2 & 177.20 & 19.84 & 18.75 \\ F7 H & 0.910 & \mybf 194.4 & 176.77 & 19.60 & 18.89 \\ F6 H & 0.963 & \mybf 198.1 & 202.97 & 17.92 & 22.56 \\ 
    \bottomrule\end{tabular}

\caption{$v_{\rm max}$ and $r_{\rm max}$ for the Aquarius halos simulated in the models $\Lambda$CDM, F6 and F7. The values for $v_{\rm max}$, $r_{\rm max}$ and $c$ are obtained with the \textsc{subfind} algorithm and neglect fifth force contributions. $c$ is the traditional concentration parameter describing the shape of the density profile.}
\label{tab:v_max}
\end{table}
\begin{figure}
\centerline{\includegraphics[width=\linewidth]{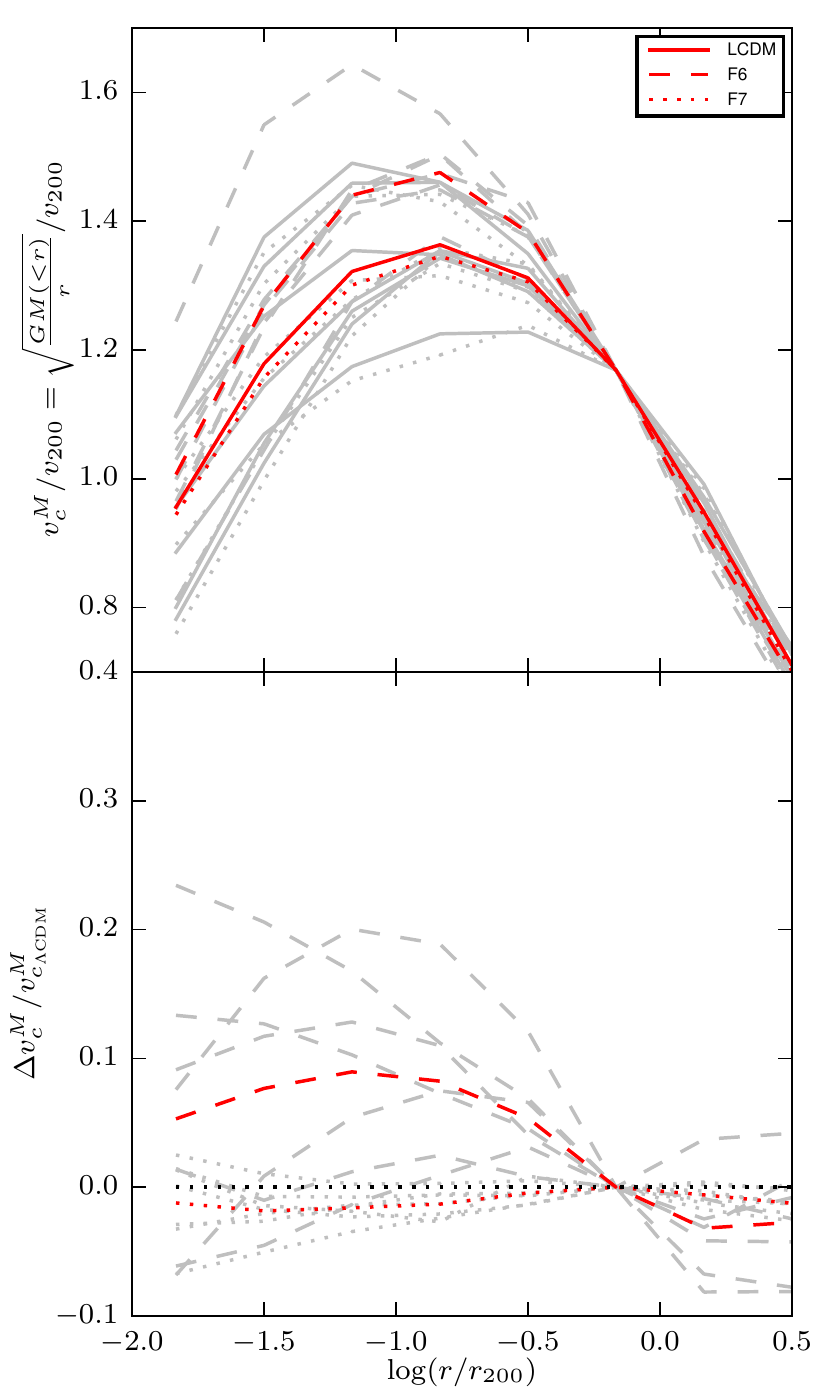}}
\caption{\textit{Upper panel:} Stacked circular velocity profiles for the simulated Aquarius halos calculated only based on the enclosed masses in units of  $v_{200}$ as a function of $r_{200}$  for $\Lambda$CDM (solid line), F6 (dashed line) and F7 (dotted line). 
\textit{Lower panel:} Relative difference in the circular velocity of the $f(R)$ models compared to $\Lambda$CDM. The \textit{dotted black line} indicates equality. The grey lines in the background show the profiles for the individual halos.}
\label{fig:velmass}
\end{figure}
\begin{figure}
\centerline{\includegraphics[width=\linewidth]{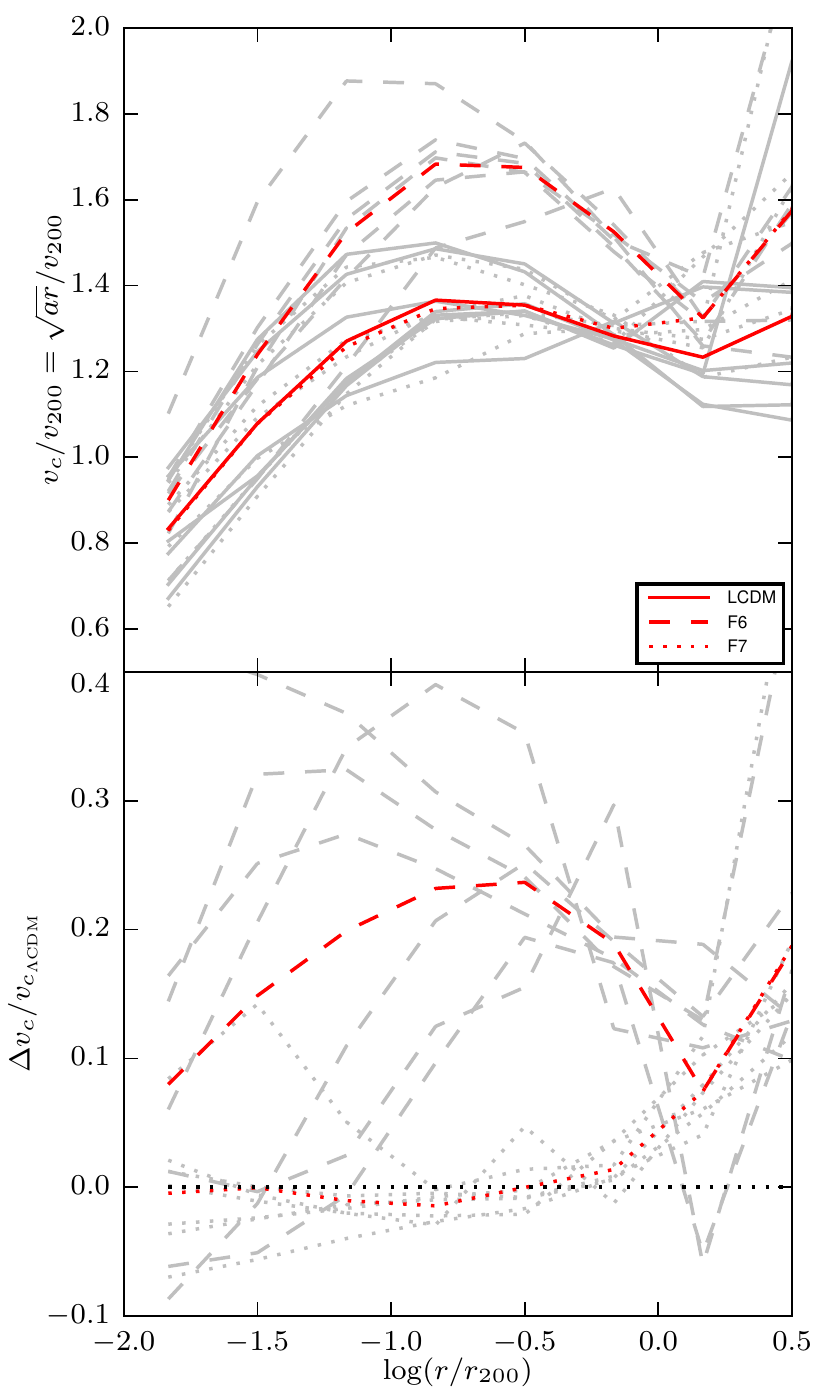}}
\caption{Same as Figure~\ref{fig:velmass} but with the circular velocities obtained from the total accelerations taking increased gravitational forces in unscreened regions for the $f(R)$ models into account.}
\label{fig:vel_acc}
\end{figure}
\begin{figure}
\centerline{\includegraphics[width=\linewidth]{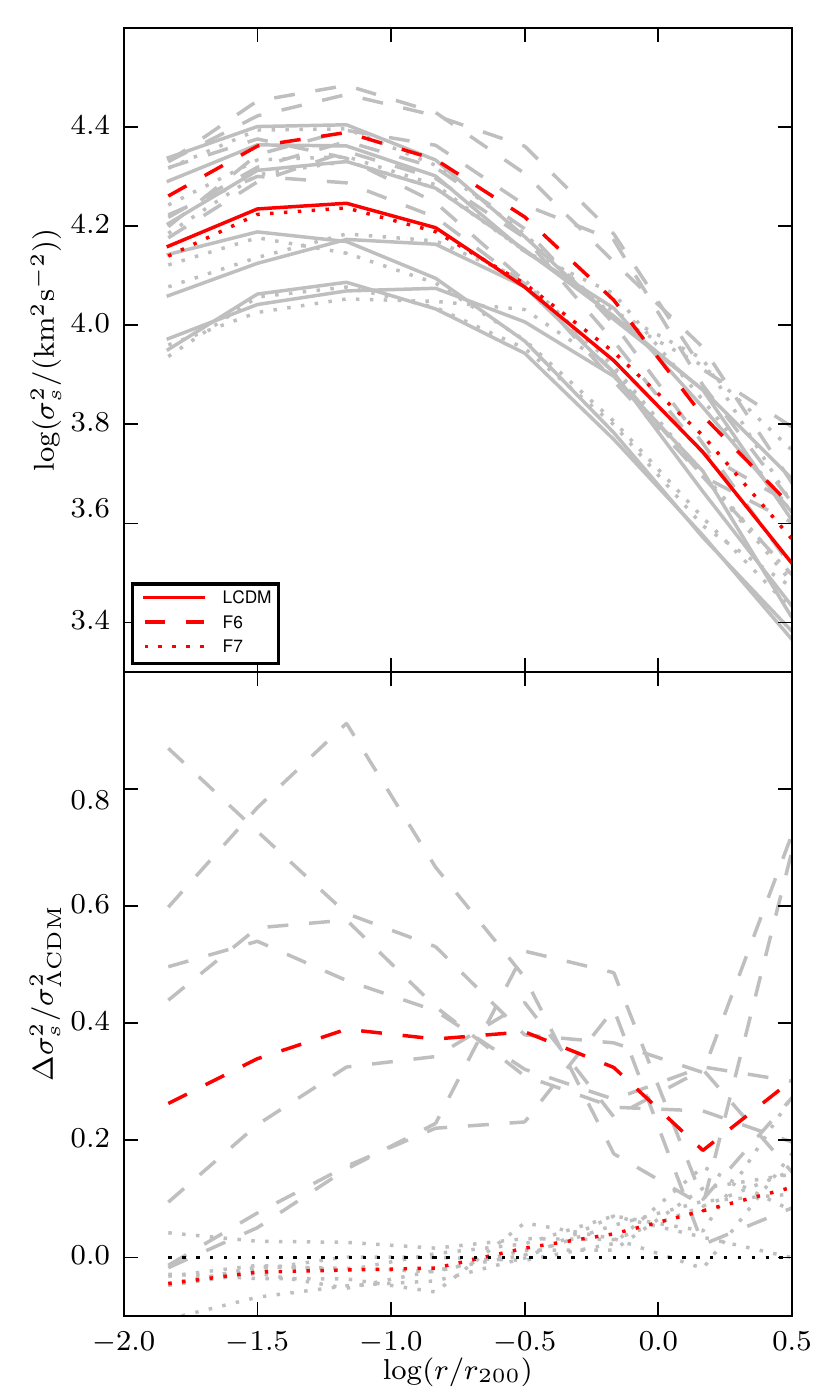}}
\caption{\textit{Upper panel:} Stacked velocity dispersion profiles for the simulated Aquarius halos for the $\Lambda$CDM (solid line), F6 (dashed line) and F7 (dotted line) cosmological models. For the $f(R)$ models the velocity dispersions were scaled according to $\sigma_s^2 = (M_{GR}/M_{f(R)})^{2/3} \sigma^2$ to filter effects due to the mass difference of the halos in the different models.
\textit{Lower panel:} Relative difference in scaled velocity dispersion between $f(R)$ gravity and $\Lambda$CDM. The black dotted line indicates equality. The grey lines in the background show the values for the individual halos.
\label{fig:vel_disp_stacked}}
\end{figure}

The ideal NFW halos analysed in the previous section have identical density profiles in the $f(R)$ and $\Lambda$CDM cosmological models. Since $f(R)$ gravity already modifies the gravitational forces during structure formation, this will in general not be the case for the outcome of self-consistent halo formation. 
Figure~\ref{fig:aqua_density} shows the density profiles of the Aquarius halos A, B and C, at $z=0$, simulated in the $f_{R0} = -10^{-6}$ (F6), $f_{R0} = -10^{-7}$ (F7) and $\Lambda$CDM cosmological models. 
The upper panel displays the density profiles relative to the critical density multiplied by $(r/r_{200})^2$.  The lower panel shows the relative difference of the density curves in $f(R)$ gravity relative to the GR runs.  
Clearly, the density profiles in F6 change significantly compared to the cosmological standard model. The density in the outer region decreases by about $10\%$ while it increases by roughly $30\%$ in the inner part. 
The transition radius depends on the mass of the halo. For F7 the changes are less significant. 
The density change in the C halo is about $10\%$ in the outer region but it is hard to tell if this is really a systematic effect or caused by small timing differences in halo assembly.

To make a robust quantitative statement about the changes in the halo densities, we stacked the density profiles of all simulated Aquarius halos. 
The profiles for the three cosmological models as well as the relative difference between $f(R)$ and $\Lambda$CDM are illustrated in Figure~\ref{fig:aqua_density_stacked}. 
The grey lines in the background show the values for the individual halos.  As already expected from the previous plot, the change in the density is quite large for the F6 model. Around $r_{200}$ the density is about $10-15\%$ lower than in a $\Lambda$CDM cosmology. At $\log_{10}(r/r_{200}) \approx -0.5$, the stacked density ratio crosses equality and reaches a maximum of about $20\%$ difference for the inner part of the objects. 
This is easily explained. The higher gravitational forces in unscreened regions in $f(R)$ gravity move mass from the outer to the inner part of the halos, thereby steepening their density profiles. 
The difference of the stacked density profiles in the $\Lambda$CDM and F7 models is consistent with zero. This shows that Milky Way sized halos are largely screened in F7.  
Keeping in mind that the F6 model passes present constraints on $f_{R0}$ {\mybf \citep{lombriser2014}}, we would like to stress that viable $f(R)$ models can hence change the density profile of Milky Way sized dark matter halos by about $20\%$. {\mybf These results are qualitatively consistent with the findings of \cite{shi2015} for Milky Way-sized halos. A direct quantitative comparison to that work is however not informative due to the much more limited resolution of the cosmological simulations employed there.}

Systematic differences in the density profiles are likely to affect the concentrations of halos. 
To investigate if the concentration shows systematic changes in $f(R)$ gravity as well, we fitted NFW-profiles \citep{navarro1997} to the density of our simulated halos for each of the three simulated models. 
Unfortunately, we found that the concentrations obtained from the fits show a relatively large residual dependence on the radial fitting range, resulting in sizable random scatter for our small halo sample. 
It is thus hard to judge on this basis if $f(R)$ gravity influences the concentration parameter in a significant way. 
As an alternative to profile fitting, we also employed another technique and obtained the concentration from the maximum of the circular velocity curve  in terms of $v_{\rm max}$ and $r_{\rm max}$ \citep{springel2008}:
\begin{align}
\delta_c &= 7.213\, \delta_V = 7.213 \times 2 \left( \frac{v_{\rm max}}{H_0\, r_{\rm max}}\right)^2, \nonumber\\ \delta_c &= \frac{200}{3} \frac{c^3}{\log(1+c) - c / (1+c)}. \label{concentration}
\end{align}
The results are summarized in Table~\ref{tab:v_max}, where $v_{\rm max}$ and $r_{\rm max}$ are obtained from the density profile directly through the \textsc{subfind} algorithm. 
These values can be used to calculate the concentration parameter of the NFW-profile. In computing $v_{\rm max}$ we ignore the force modifications which occur in unscreened regions in $f(R)$-gravity, i.e.~$v_{\rm max}$ and $r_{\rm max}$ are completely determined by the density profile, as appropriate for measuring its concentration. They should not be confused with the velocities shown in Figure~\ref{fig:vel_acc}. The numbers in Table~\ref{tab:v_max} are rather connected to the curves in Figure~\ref{fig:velmass}.

Comparing the concentration $c$ of the objects simulated in F6 and GR, we find that the concentrations are increased for the B, D, E G and H halo. For the A and the C halo, however, the concentration parameter decreases in $f(R)$ gravity compared to $\Lambda$CDM. 
One can thus conclude that there appears to be a slight trend to higher concentrations in $f(R)$ gravity, but a much larger number of halos would be needed to establish this finding robustly. 
It would also be important to carefully select sufficiently relaxed halos \citep[e.g. as in][]{neto2007} to avoid influences from mergers or large substructures. Because the effects in F7 are weaker, it is even harder to demonstrate if and how the concentration changes for this model.

Figure~\ref{fig:velmass} shows stacked circular velocity profiles for the Aquarius halos in F6, F7 and GR as well as the relative differences between the modified gravity models and $\Lambda$CDM. 
The grey lines in the background display the velocity profiles of the individual halos. For this plot, the velocities were obtained from the enclosed masses using the standard relation for Newtonian gravity, hence neglecting any $f(R)$ effects other than those encoded in the mass distribution. In order to clearly separate effects which are induced by the slightly higher masses in $f(R)$ gravity, the velocities are scaled with $v_{200} \equiv ({G M_{200}}/{r_{200}})^{1/2}$. 
The velocity curves are therefore a direct measure of the mass profile and useful to determine, for example, the concentration of the mass profile in the standard way. 
It is not surprising that the relative difference between the F6 model and GR is of order $10\%$ compared to a $20\%$ difference in density in the inner part of the halo ($v \propto \sqrt{M}$). 
In contrast to the density, the velocity does not drop significantly below the $\Lambda$CDM value in the outer regions since the velocities see the cumulative mass profile which includes the higher density in the center. 
The slightly lower values outside of $r_{200}$ are due to the rescaling with $v_{200}$. Since the density profile does not change noticeably in F7 the change in circular velocities is negligible as well.

As a cautionary remark we would like to add that the velocities shown in Figure~\ref{fig:velmass} should not be confused with observable circular velocities. 
For those, the differences in the accelerations between the different models must be included in the analysis. This was done for Figure~\ref{fig:vel_acc}, where we show stacked circular velocity profiles obtained from the total accelerations. In the upper panel, the absolute values of the velocities are displayed for the three simulated models, the lower panel shows the relative differences of $f(R)$ gravity to GR. 
The velocities in the F6 model are significantly higher compared to standard gravity and to the previous plot.  This is easily explained. In addition to the higher densities in the inner region of the halos, higher gravitational accelerations in unscreened regions force the DM particles in the simulation to orbit faster in order to prevent infall. 
As a result, the circular velocities are increased by up to $25\%$ compared to GR in unscreened regions.  
Although a $25\%$ difference in the velocity profile for an allowed $f(R)$ model seems large, one has to keep in mind that the effects will be at least partially degenerate with (the quite uncertain) baryonic physics \citep{vogelsberger2014, marinacci2014} and uncertainty in the halo mass. Also, the error bars of the current observational constraints \citep{avila2008, hall2012, mcgaugh2012} allow a broad range of velocities. It will therefore be hard to constrain $f_{R0}$ relying on the rotation curves of Milky Way sized objects.  

For the F7 model, the circular velocities stay unchanged in the inner region of the halo because the $f(R)$ modifications to gravity are screened by the chameleon mechanism. 
At around $0.5\, r_{200}$ the relative velocity difference increases and reaches $20\%$ at $3\, r_{200}$. This can be explained by two mechanisms. 
On the one hand, the halo becomes unscreened in the outer parts due to the shallower gravitational potential. The gravitational forces are thus by a factor of $4/3$ higher and increase the velocities even if the density is the same. 
On the other hand, all velocities increase outside $r_{200}$ because the particles start to see other objects and are not in virial equilibrium. In combination with higher forces, this adds another boost to the velocities.
\begin{figure} \centerline{\includegraphics[width=\linewidth]{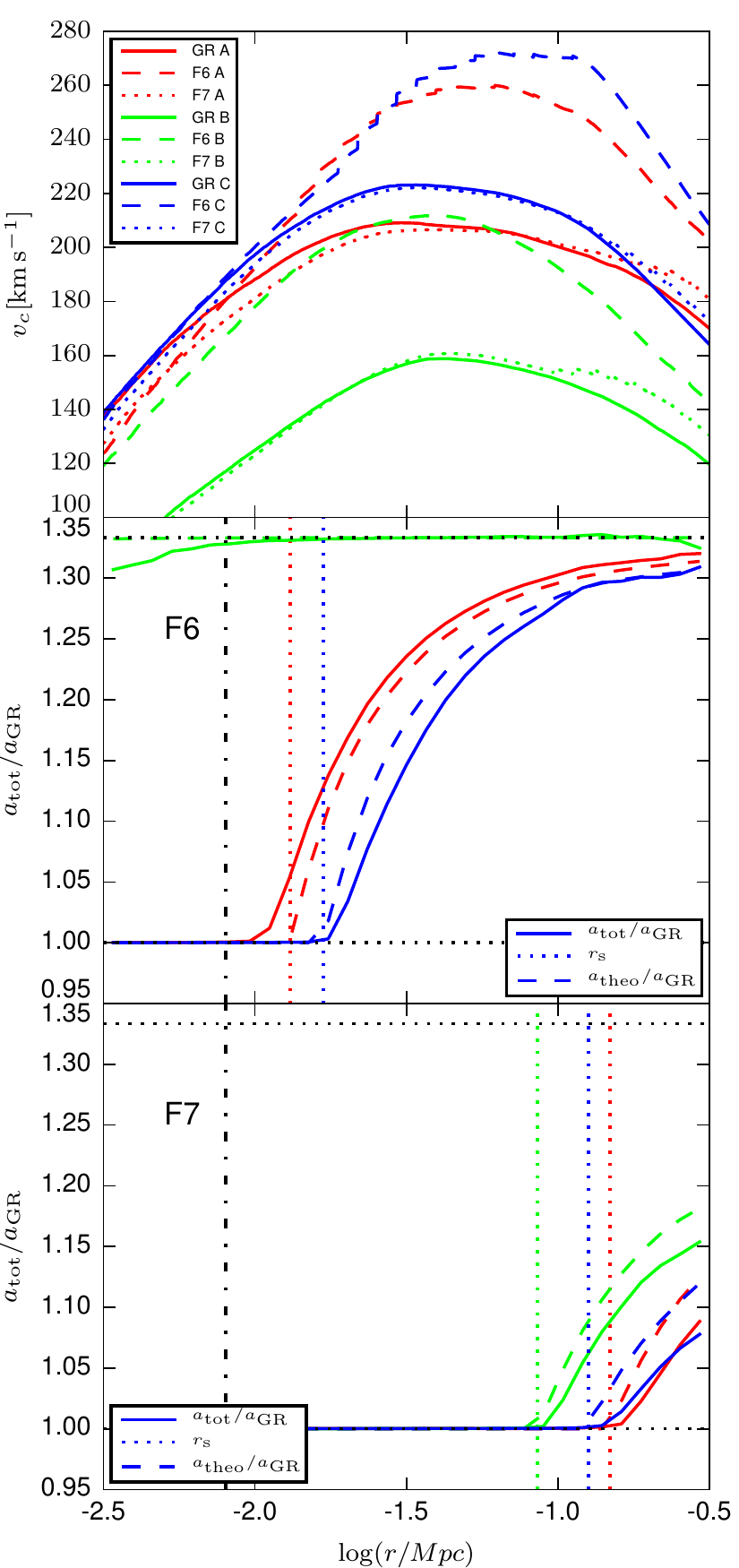}} \caption{Circular velocity and acceleration profiles for the halos A (red lines), B (green lines) and C (blue lines). 
\textit{Upper panel:} Circular velocity profiles for $\Lambda$CDM (solid lines), F6 (dashed lines) and F7 (dotted lines) taking the increased accelerations due to modified gravity fifth forces in the $f(R)$ models into account. 
\textit{Center panel:} Ratio of the total force to GR force for the three halos in the F6 cosmology. The results from the simulations are shown as solid lines. Dashed lines indicate the theoretical expectations. The predicted screening radii, $r_s$, are shown as vertical dotted lines. The two black horizontal dotted lines indicate equality and the maximum value for the force ratio of $a_{\rm tot}/a_{\rm GR} = 4/3$. 
For reference, the distance of the Sun from the Galactic centre is indicated by the vertical dashed dotted line. \textit{Lower panel:} Same as the center panel but for the F7 cosmology.}
\label{fig:aqua_acc}
\end{figure}

Figure~\ref{fig:vel_disp_stacked} shows the stacked velocity dispersion profiles for F6, F7 and GR (upper panel) as well as the relative difference of the modified gravity values to $\Lambda$CDM (lower panel). 
Since the difference in halo mass between the models for a given object (see Table \ref{tab:v_max}) would also lead to differences in the velocity dispersion, we scaled it according to $\sigma_s^2 = (M_{\rm GR}/M_{f(R)})^{2/3} \sigma^2$ for the $f(R)$ curves to account for the mass difference. The scaled velocity dispersion shown in the plot is therefore a measure {\mybf of} how the velocity dispersions of halos of a given mass would change in $f(R)$-gravity.
For the F6 run, we find velocity dispersions increased by about $40\%$ in the inner part ($-1.5 < \log_{10}(r/r_{200}) < -0.5$) which is again a result of the higher densities in this central part of the halo and the increased gravitational forces.  
In the outer regions, the cumulative mass profile stays unchanged compared to GR and thus only the $4/3$ enhancement of the forces contributes to the about $30\%$ higher velocity dispersion. Outside $r_{200}$, the halo shows again larger differences between the models due to a lack of virialisation.

The $40\%$ difference between the $f(R)$ and $\Lambda$CDM cosmological models is slightly higher than the values for unscreened halos of about $30\%$ reported in \cite{schmidt2010}, \cite{lam2012}, \cite{arnold2014}, \cite{gronke2015} and {\mybf\cite{shi2015}}. There are several reasons for this difference. First, all of these other works used cosmological simulations with mass resolutions poorer by factors $10-100$ (relative to the mass of the considered object) compared to the high resolution simulations in this work. They were therefore most likely not capable of capturing the increased density in full in the inner region of the halos. 
Second, the previous works either present the averaged velocity dispersion of the whole object or do not show the profiles in the inner part. 
For both cases, the velocities will be dominated by the outer regions which obey a smaller velocity dispersion.  
We note that for the weaker F7 model, the velocity dispersion stays unchanged in the central region because the fifth force is again screened. Further out, the difference to GR grows to $10\%$ at $r_{200}$.

In the following we like to extend our comparison of the theoretically predicted screening radius and fifth force (see section \ref{subsec:spherical}) to the simulated Aquarius halos. 
In contrast to perfectly symmetric NFW profiles the simulated halos are ellipsoidal and feature substructures which breaks spherical symmetry. 
Our goal is to find out if the theoretical approximations are nevertheless applicable and reasonably accurate for realistic halos.  
The upper panel of Figure~\ref{fig:aqua_acc} shows the circular velocity profiles of the Aquarius halos A, B and C, for F6, F7 and GR. The profiles are, as in Figure~\ref{fig:NFW}, obtained from the enclosed mass with an additional factor for the increased forces in $f(R)$-gravity. The small steps visible in some of the $f(R)$ profiles are due to the binning of the acceleration ratio. For the A and the C halo, the velocities are increased by about $20 - 30\%$ with respect to GR in the outer region. Moving further in, the difference between the $f(R)$ and $\Lambda$CDM curves decreases due to chameleon screening until they almost match. 
The B halo has a slightly lower mass. Its velocity curve is by $20 - 30\%$ higher than the curve obtained from the GR simulation over the whole range of radii shown in the plot. This suggests that this halo is largely unscreened.

These results are confirmed by the acceleration ratios for the F6 model displayed in the middle panel. For the two massive halos, the acceleration ratio drops to unity at $r\approx 0.02\, {\rm Mpc}$. Inside this radius, the $f(R)$ modifications to gravity are screened by the chameleon mechanism. 
For the B halo, the acceleration ratio stays at the theoretical maximum of $1.33$ over almost the whole range shown in the plot. 
Only in the innermost part there is a slight deviation which could naively be interpreted as the onset of screening, but is more likely an effect caused by the lack of resolution of the AMR-grid in the central region of the least massive object.

In the weaker F7 model, all three objects are almost totally screened. 
The velocity profiles coincide with the $\Lambda$CDM curves. Only in the very outer region, chameleon screening breaks down and the velocities in F7 are increased with respect to GR. 
 Again, the acceleration ratios confirm this result. The lower panel of Figure~\ref{fig:aqua_acc} shows that the fifth force vanishes everywhere, except in the outskirts.

The middle and the lower panel of Figure~\ref{fig:aqua_acc} also display the theoretically expected screening radius and force ratio. 
It turns out that the analytical screening radius $r_s$ (again, calculated from Eqn.~\ref{theo_rs}) is still a very good proxy for the radius where the actual force ratio drops to unity, although it is unsurprisingly not as accurate as for the ideal NFW profiles (Figure~\ref{fig:NFW}). 
The force ratios are also captured pretty well by the theoretical predictions, but are only accurate to about $5\%$ for realistic halos. 
As already mentioned, these differences occur due to the asymmetric shapes of the simulated halos and the presences of substructures. In the vicinity of a large subhalo the main halo may already be screened while the chameleon screening has not necessarily set in at the same radial distance on the opposite side of the halo. 
So our results show that the analytic model predictions are quite powerful for reasonably smooth halos whereas for objects with a high abundance of massive substructures, such as forming galaxy clusters or groups, their accuracy is somewhat compromised. 
This then also underlines that for scenarios with a very non-linear dependence of the fifth force on the density field, numerical simulations are essential to accurately capture all relevant effects.

Coming back to the requirement that the Solar system should be screened within the Milky Way, it is evident from Figure~\ref{fig:aqua_acc} that even the stronger F6 model fulfills this constraint. 
For the two more massive objects A and C, which are closer to the Milky Way in mass, the halo is already completely screened at the radius of the Solar system, i.e. $r\approx 8\,{\rm kpc}$. 
There is nevertheless not much space for more strongly modified models. This finding is consistent with previous constraints of the $f(R)$-model \citep[see e.g.][]{terukina2014}.

\section{Summary and Conclusions}
\label{sec:conclusions}

We analysed the properties of Milky Way sized dark matter halos in \cite{husa2007} $f(R)$-gravity employing cosmological zoom simulations. Using our simulation code \textsc{mg-gadget}, we simulated a set of 7 DM halos from the Aquarius suite in the F6 and F7 models, as well as in the $\Lambda$CDM cosmology, for comparison.  We also compared the simulation results against an analytical estimate of the fifth force in DM halos.  Our main findings can be summarized as follows.

\begin{itemize}
\item The theoretical predictions for the screening radius and the fifth force inside a spherical object derived in \cite{vikram2014} (see also Section \ref{subsec:spherical}) reproduce the results obtained with our numerical modified gravity solver to high precision for ideal NFW-halos. For realistic halos from the cosmological simulations, the applicability is somewhat limited due to triaxial halo shapes and substructures. The theoretical estimate can nevertheless serve as a proxy for reasonably smooth and relaxed halos in relatively isolated environments.
\item The self-similarity of DM halos observed in $\Lambda$CDM is broken in $f(R)$-gravity due to the scale introduced by the screening radius for a given choice of $f_{R0}$. It can be approximately restored by appropriately scaling both $f_{R0}$ and the mass of the object.
\item Our simulations show that the density of a Milky Way sized halo in F6 modified gravity is increased in the inner part, while it is slightly lower around $r_{200}$ compared to GR. For the F7 model, the density profiles are largely unchanged. 
\item The impact of $f(R)$-gravity on the mean halo concentration parameter cannot be reliably quantified from our simulations due to random scatter in fitting individual NFW density profiles and the small sample size.  The density profiles, nevertheless, suggest a higher concentration of DM halos in $f(R)$-gravity compared to $\Lambda$CDM {\mybf \citep[as previously reported in][]{shi2015}}. 
As higher concentrations imply a smaller Milky Way mass to match observational constraints, this appears to provide yet another potential solution for the too-big-to-fail problem \citep{Boylan-Kolchin2011, Cautun2014}.
\item Circular velocities in $f(R)$ gravity are increased in unscreened regions with respect to the $\Lambda$CDM cosmology. Velocities calculated in the standard way only from the enclosed mass show a relative enhancement of about $12\%$ in F6, while there is almost no difference for the F7 model due to the screening mechanism. 
For the circular velocities calculated more appropriately from the accelerations, there is an additional boost from the increased forces resulting in up to a $30\%$ difference relative to GR for the F6 model, and in about $10\%$ higher velocities for the F7 model in the unscreened outer parts of the halos. 
One should pay attention that these two measures, which are equivalent in a $\Lambda$CDM cosmology, yield different results in $f(R)$-gravity.
\item The velocity dispersion inside the halos is increased by up to $40\%$ in F6 with respect to standard gravity. This relative difference is larger than the enhancement of about $30\%$ which is found in previous works. 
We conclude that earlier works most likely did not have enough mass resolution to safely capture the effects on the density profile during structure formation and therefore missed an imported contribution to the enhanced velocity dispersion. 
{\mybf Although the size of our sample is limited and the scatter between the halos is quite large, we think that this result is still reliable. The scatter is mainly induced by the screening mechanism due to the different halo masses. Completely unscreened halos have an even larger difference in velocity dispersion in the central region. Regardless of its small size, the sample is mass-selected to match the mass of the Milky Way. In addition, the velocity dispersions in simulations carried out with \textsc{mg-gadget} agree with results from other codes when using the same resolution \citep[see][]{winther2015}.} 
For the F7 model, the differences to GR are much weaker due to the chameleon screening mechanism. 
\item The simulations show that the ratio of total-to-GR acceleration is increased by the theoretically expected factor of $4/3$ in the outer parts of the halos for F6 gravity. In the inner parts, the more massive halos of our sample are screened and thus show no difference in the force compared to GR. In the F7 model, the halos are almost completely screened and exhibit only a small force difference around $r_{200}$. 
\item The halos which have a mass close to that of the Milky Way are completely screened at the position of the Solar system both in the F6 and the F7 model. Halos with slightly lower mass do not show screening at the Solar circle, underlining that F6 is the strongest allowed $f(R)$ model. This is consistent with Solar system constraints on $\bar{f}_{R0}$ from the literature.  
\end{itemize}

All in all we conclude that the effect of viable $f(R)$-gravity models on the density profiles and velocity dispersions of Milky Way like halos are quite large. Both simulated parameter values of the \cite{husa2007} model, F7 and F6, are, according to our simulations, fully consistent with local constraints. 
Even models which are screened at the galactocentric radius of the Solar system can exhibit large differences in the velocity dispersion and the density profile. In the context of upcoming missions which are designed to test gravity on large scales, it is therefore essential to explore the alternatives to GR and the cosmological standard model, $\Lambda$CDM, in detail in order to provide reliable information on the effect of these theories on cosmological observables.

\section*{Acknowledgements}

VS and CA would like to thank the Klaus Tschira foundation and
acknowledge support from the Deutsche Forschungsgemeinschaft (DFG)
through Transregio 33, ``The Dark Universe''.  EP gratefully
acknowledges support by the Kavli Foundation and the FP7 ERC Advanced
Grant Emergence-320596. The authors acknowledge CPU-time from the
Juelich Supercomputer Centre on the JURECA system.


\bibliographystyle{mnras}
\bibliography{paper}


\end{document}

%% file: JournalAbbr.tex
\def\apj{ApJ}%
\def\apjl{ApJ}%
\def\apjs{ApJS}%
\def\ao{Appl.~Opt.}%
\def\aap{A\&A}%
\def\mnras{MNRAS}%
\def\prd{Phys.~Rev.~D}%
\def\physrep{Phys.~Rep.}%
\def\jcap{JCAP}